\documentclass[12pt]{article}
\usepackage{amssymb,amsmath,amsthm,amsfonts,latexsym}
\usepackage{mathptmx}       
\usepackage{helvet}         
\usepackage{courier}        
\usepackage{type1cm}        
\usepackage{makeidx}         
\usepackage{graphicx}        
\usepackage{multicol}        
\usepackage[bottom]{footmisc}

\makeindex             

\newcommand{\be}{\begin{equation}}
\newcommand{\ee}{\end{equation}}

\newcommand{\ds}{\displaystyle}

\def\E{\mathbb{E}}
\def\I{\mathbb{I}}
\def\J{\mathbb{J}}
\def\N{\mathbb{N}}
\def\R{\mathbb{R}}
\def\S{\mathbb{S}}
\def\Z{\mathbb{Z}}
\def\a{\alpha}

\def\a{\alpha }

\begin{document}

\begin{center}
\bf{\LARGE 
Group theoretical aspects of $L^2(\R^+)$, $L^2(\R^2)$ and associated Laguerre polynomials}
\footnote{Contribution to the 31st International Colloquium on Group Theoretical Methods in Physics, Rio de Janeiro, June 19-25, 2016. Accepted in {\sl Springer  Proceeding Series}.}
\end{center}

\bigskip\bigskip

\begin{center}
E. Celeghini$^{1,2}$, M.A. del Olmo$^2$.
\end{center}

\begin{center}
$^1${\sl Dipartimento di Fisica, Universit\`a  di Firenze and
INFN--Sezione di
Firenze \\
I50019 Sesto Fiorentino,  Firenze, Italy}\\
\medskip

$^2${\sl Departamento de F\'{\i}sica Te\'orica and IMUVA, Universidad de
Valladolid, \\
E-47011, Valladolid, Spain.}\\
\medskip

{e-mail: celeghini@fi.infn.it, olmo@fta.uva.es}

\end{center}

\abstract{
A ladder algebraic structure for $L^2(\R^+)$  which closes the Lie algebra $h(1)\oplus h(1)$, where $h(1)$ is the Heisenberg-Weyl algebra,  is presented in terms of a basis of associated  Laguerre polynomials. Using the  Schwinger method  the quadratic generators   that span   the alternative Lie algebras $so(3)$, \\
$so(2,1)$ and $so(3,2)$ are also constructed. These  families of (pseudo) orthogonal algebras  also allow to obtain unitary irreducible representations  in $L^2(\R^2)$ similar to those of the spherical harmonics.
}

\section{Introduction}
\label{sec:1}

The associated Laguerre polynomials (ALP) \cite{szego}, $L_n^{(\alpha)}(x)$ 	$\left( x\in [0 , \infty \right)$,
$n=0,1,2,\cdots$ and $\alpha$    real fixed parameter, continuous and $>  -1$),
are defined by the 2nd order differential equation (DE)
\be\label{defLag}
\left[x \frac{d^2}{dx^2} + (1+\alpha-x) \frac{d}{dx} + n \right] L^{(\alpha)}_n(x) =0 \;.
\ee
The ALPs reduce to the
Laguerre polynomials for\, $\alpha = 0$.
From the many recurrence relations that they verify \cite{szego,NITS,abramowitz},
we start from the following ones
\be\label{Moperators}
\left[-\frac{d}{dx}  + 1 \right] L^{(\alpha)}_n(x) = L^{(\alpha+1)}_n(x)\,,
\quad
\left[x \frac{d}{dx} + \alpha \right] L^{(\alpha)}_n(x) = (n + \alpha )\,
L^{(\alpha -1)}_n(x)\,.
\ee

For $\alpha >-1$ and
fixed, the ALP  $L_n^{(\alpha)}(r)$ are orthogonal
in  the label $n$ with respect the weight measure
$d\mu(x)=x^{\a}\, e^{-x}\, dx$
\[
\int_0^\infty\; dx \;x^\a\; e^{-x}\, L_n^{(\alpha)}(x)\; L_{n'}^{(\alpha)}(x)\; =\;
\frac{\Gamma(n+\a+1)}{n!}\;  \delta_{n n'}
\, .
\]
For $\alpha$ integer such that
\; $0 \le \alpha  \le n$\,, we have the generalization \cite{szego}
$$
L_n^{(-\alpha)}(x) := \frac{\Gamma(n-\alpha+1)}{\Gamma(n+1)} \,(-x)^\alpha \, L_{n-\alpha}^{(\alpha)}(x)\, .
$$
Hereafter we assume here
$n \in \N ,\,\, \alpha \in \Z \,,\,\,
n-\alpha \in \N$ and
we consider\, $\alpha$\, as a label, like $n$, and not a parameter
fixed at the beginning.

Following the approach of previous works \cite{olmo13,olmo13a,olmo15,olmo15a}  we introduce
now a set of alternative functions
 including also the weight measure, in such a way to obtain the orthonormal
bases we   are used to in Quantum Mechanics
\[
M^{(\alpha)}_n(x) := \sqrt{\frac{\Gamma(n+1)}{\Gamma(n+\alpha+1)}}\; x^{\alpha/2}\;
e^{-x/2}\;  L^{(\alpha)}_n(x) \,.
\]
For each fixed value of $\alpha \geq -n$ and  $n \in \N$, the set of
$M_n^{(\alpha)}(x)$,  is a
basis of $L^2(\R^+)$
\[ \label{MnaOrthNorm}
 \int_{0}^{\infty} M^{(\alpha)}_{n}(x)\, \, M^{(\alpha)}_{m}(x) \;dx =
 \delta_{n m}\,,\qquad
\displaystyle \sum_{n=0}^{\infty}  \;M^{(\alpha)}_{n}(x)\,\, M^{(\alpha)}_{n}(x') = \delta(x-x') \,.
\]

\section{The symmetry algebra
$h(1)_n \oplus h(1)_p$}
\label{sec:2}

The eqs. \eqref{Moperators} rewritten in terms of $M_n^{(\alpha)}$ take the form
\be
\begin{array}{lll}\label{recurr5}
\displaystyle
\left[-\sqrt{x} \frac{d}{dx}+\frac{1}{2\sqrt{x}}(\alpha + x)\right] \;M^{(\alpha)}_n(x)\; &=&\;
\sqrt{n+\alpha+1} \;M^{(\alpha+1)}_n(x)\, ,
\\[0.3cm]
\displaystyle
\left[\sqrt{x} \frac{d}{dx} + \frac{1}{2\sqrt{x}}(\alpha + x)\right] \;M^{(\alpha)}_n(x)\;& =&\;
\sqrt{n+\alpha} \;M^{(\alpha-1)}_n(x) \,,
\end{array}
\ee
where  $p :=n + \alpha$ plays, for $n$ fixed, the role of eigenvalue of the number operator
in a  Heisenberg-Weyl algebra, $h(1)$, realized on the space of functions $M^{\alpha}_n(x)$. It is indeed a positive integer like $n$, so that
we can define the new functions
${{\cal M}_{n,p}(x) := M_n^{(p-n)}(x)}$, that by inspection
 are symmetric in the interchange
 $n \Leftrightarrow p$, i.e.
${\cal M}_{n,p}(x) \,=\,  (-1)^{p-n} \,{\cal M}_{p,n}(x)\,.$
The previous recurrence relations \eqref{recurr5}
can thus be rewritten
\be
\begin{array}{lll}\label{recurr6}
\displaystyle
\left[-\sqrt{x} \frac{d}{dx}+ \frac{\sqrt{x}}{2} +  \frac{p-n}{2\sqrt{x}}\right]\; {\cal M}_{n,p}(x) \;&=&\;
\sqrt{p+1} \;{\cal M}_{n,p+1}(x)\,,
\\[0.5cm]
\displaystyle
\left[\sqrt{x} \frac{d}{dx} + \frac{\sqrt{x}}{2} + \frac{p-n}{2\sqrt{x}}\right]\; {\cal M}_{n,p}(x) \;&=&\;
\sqrt{p} \;{\cal M}_{n,p-1}(x) \,.
\end{array}
\ee

To construct the operatorial structure corresponding to the recurrence relations   we define now
   four  operators $X$, $D_x$, $N$ and $P$
\[\begin{array}{lllll}
\displaystyle
X\; {\cal M}_{n,p}(x)\; &=&\; x\; {\cal M}_{n,p}(x)\,,\qquad
D_x\; {\cal M}_{n,p}(x)\; &=&\; \frac{d{\cal M}_{n,p}(x)}{dx} \,,
\\[0.3cm]
N\; {\cal M}_{n,p}(x)\; &=&\; n\; {\cal M}_{n,p}(x)\,,\qquad
P\; {\cal M}_{n,p}(x)\; &=&\; p\; {\cal M}_{n,p}(x) \,.
\end{array}\]
Then, the 2nd order DE  \eqref{defLag} becomes
\be\label{defLag2}
\E\;{\cal M}_{n,p}(x) =0\,,
\ee
where
\[
\E\,:=\,X D_x^2 + D_x + \frac{N + P +1}{2}  -\frac{1}{4 X} (P - N)^2 -\frac{X}{4}\,.
\]

Moreover from  \eqref{recurr6} we get  the differential operators (DOs)
\be\label{oper1}
{\bold b}^\pm: =  \mp\sqrt{X} D_x +\frac{\sqrt{X}}{2} + \frac{1}{2\sqrt{X}}(P-N)\, ,
\ee
that act on  the functions ${\cal M}_{n,p}(x)$  in such a way that $\Delta n = 0$ and $\Delta p = \pm 1$.
Since 
$
[ {\bold b}^-, {\bold b}^+ ] =  \I\,$ they close an $h(1)$ algebra,
($h(1)_{p}$) with 
 quadratic Casimir   
$
{\cal C}_p=\{{\bold b}^-,{\bold b}^+\}-2(P+1/2)
$
  verifying
$
{\cal C}_p\, {\cal M}_{n,p}(x)= -2 \, \E \,{\cal M}_{n,p}(x)=0\,.
$

Now taking into account the symmetry under the interchange
$n \Leftrightarrow p$ of  ${\cal M}_{n,p}(x)$
we can define the operators
${\bold a}^\pm(N,P):=-{\bold b}^\pm(P,N)$
 that  change
the labels  of $ {\cal M}_{n,p}(x)$ as $\Delta p = 0$ and $\Delta n = \pm 1$.
 Their explicit action on  ${\cal M}_{n,p}(x)$ is indeed
\[
{\bold a}^+ \; {\cal M}_{n,p}(x)\, =\, \sqrt{n+1}\; {\cal M}_{n+1,p}(x)\,,\qquad
{\bold a}^- \; {\cal M}_{n,p}(x)\, =\, \sqrt{n}\; {\cal M}_{n-1,p}(x) \,.
\]
The two operators ${\bold a}^\pm $ determine thus another HW algebra,   $h(1)_{n}$.
Since these bosonic operators ${\bold a}^\pm$ and $ {\bold b}^\pm$ commute among them we have obtained in this way the  global
algebra $h(1)_n \oplus h(1)_p$.

Moreover inside the Universal Enveloping Algebra $UEA \left[h(1)_n \oplus h(1)_p\right]$ other algebras preserving the parity of $n+p$ can be found
by the Schwinger procedure \cite{schwinger} as we will do in the next section.

\section{$so(3)$,  $so(2,1)$ and $so(3,2)$ symmetries}
\label{sec:2}
\subsubsection*{$so(3)$  symmetry}

We start from
$J_\pm:= {\bold a}_\pm\,{\bold b}_\mp$
obtaining  2nd order DOs that,
taking into account  eq. \eqref{defLag2}, can be rewritten in the space $\{{\cal M}_{n,p}(x)\}$
 as  1st order DOs
\be\label{jotaoperators}
J_\pm=\mp\,D_x\,(N-P\pm 1)+\ds\frac{1}{2 X}\,
(N-P\pm 1)(N-P)
-\frac{1}{2}(N+P+1)\, .
\ee
Defining
$J_3:=({\bold a}_-\,{\bold a}_+ - {\bold b}_- \,{\bold b}_+)/2\;\equiv \;{(N-P)}/{2}$
 we  see that $\{J_\pm,J_3\}$ close a $su(2)$ algebra in the space $\{{\cal M}_{n,p}(x)\}$ since
$[J_+,J_-]= 2J_3- \frac{8}{X}\, J_3\, \E .$
The action of $J_\pm$  is
\[
J_+
 \;{\cal M}_{n,p}(x) = \sqrt{(n+1)\, p}\; {\cal M}_{n+1,p-1}(x)\, ,\quad
 J_-
 \;{\cal M}_{n,p}(x) = \sqrt{n\, (p+1)}\; {\cal M}_{n-1,p+1}(x)\, .
\]
Also the Casimir of $su(2)$, ${\cal{C}}_{su(2)}=J_3^2+\frac12\{ J_+,J_-\}$ is closely
related to eq.  \eqref{defLag2} as
${\cal{C}}_{su(2)}=J(J+1)+\frac{1}{X}\,
(4 J_3^2+1)\,  \E$,
where $J$ is the diagonal operator
$J:=(N+P)/2 $.


\subsubsection*{$so(2,1)$ symmetry}

In a similar way   we can define  the operators $K_\pm:= {\bold a}_\pm\,{\bold b}_\pm$,
  such that, like in the case of the operators $J_\pm$, we find in the space $\{{\cal M}_{n,p}(x)\}$
\be\label{kaoperators}
K_+=X\,D_x+\ds\frac{1}{2}(N+P+2-X)\,,\qquad
K_-=-X\,D_x+\ds \frac{1}{2}(N+P-X)\,.
\ee
Both operators together with
$
K_3:=({\bold a}_-\,{\bold a}_+ + {\bold b}_+ \,{\bold b}_-)/2\;\equiv\;{(N+P+1)}/{2}$
 determine a $su(1,1)$ algebra
$$
[K_3,K_\pm]=\pm   K_\pm\,,\qquad [K_+,K_-]=- 2 K_3\,,
$$
since the action  on the functions ${\cal M}_{n,p}(x)$ is
\[
K_+ \;{\cal M}_{n,p}(x) = \sqrt{(n+1)(p+1)}\; {\cal M}_{n+1,p+1}(x) \,,\quad
K_- \;{\cal M}_{n,p}(x) = \sqrt{n\, p}\; {\cal M}_{n-1,p-1}(x)\,.
\]
The Casimir of $su(1,1)$, ${\cal{C}}_{su(1,1)}=K_3^2-\frac12\{ K_+,K_-\}$, is also connected with eq. \eqref{defLag2}
 as
$
{\cal{C}}_{su(1,1)}=(M^2-\frac14)+ {X}\,  \E
$,
where  $M=J_3 :={(N-P)}/{2}$.

\subsubsection*{More $so(2,1)$ symmetries}

The commutators of  $J_\pm$ and $K_\pm$ give the new operators
$$
R_\pm:=\pm [J_\pm,K_\pm]\,,
\qquad
S_\pm:=\pm [J_\mp,K_\pm]\,.
$$
Provided that we define
$R_3:=J+M+1/2$ and $S_3:=J-M+1/2$, they close two $so(2,1)$ algebras with
commutators
$$
[R_+,R_-]=-4 R_3\,,
\qquad [R_3,R_\pm]=\pm 2 R_\pm\,,
$$
and  Casimir $
{\cal C}_R=R_3^2-\frac12\{R_+,R_-\}\,=\, -\frac34 +
\frac{1}{X}\,(1+(X+2 M)^2) \, \E$ and similarly for $\{S_\pm,S_3\}$.
Note that under the interchange
$m\leftrightarrow -m$ we have
 $\{R_\pm,R_3\} \leftrightarrow  \{S_\pm,S_3\}$.


\subsubsection*{$so(3,2)$ symmetry}

All the operators  $\{K_\pm, L_\pm,R_\pm, S_\pm, J, M\}$ can be written on the space
$\{{\cal M}_{n,p}(x)\}$
as 1st order DOs. All together they determine on $\{{\cal M}_{n,p}(x)\}$ the representation  of  the Lie algebra $so(3,2)$  with $C^{so(3,2)}_2 =- 5/4 $ .

\section{Representations of $so(3)$, $so(2,1)$ and $so(3,2)$ on the plane}

We introduce   now the operators directly related to $so(3)$,
$J := (N+P)/2$ and $J_3\equiv M := (N-P)/2$,
and define
\[
{\cal L}^m_j(x) :={\cal M}_{j+m, j-m}(x)
=\sqrt{\frac{(j+m)!}{(j-m)!}}
\,x^{-m}\,e^{-x/2}\,L_{j+m}^{(-2m)}(x)\,.
\]

 The operators  $J_3$ and $J_  \pm$  \eqref{jotaoperators},
  rewritten in terms of $J$ and $M$,
act on $\{{\cal L}^m_j(x)\}$ as
\[
J_3 \;{\cal L}^m_j(x) = m\; {\cal L}^m_{j}(x)\, , \quad
J_\pm \;{\cal L}^m_j(x)  = \sqrt{(j\mp m)(j\pm m+1)}\; {\cal L}^{m\pm 1}_j(x)\,.
\]
So, $\{{\cal L}^m_j(x)\}$ with $ j\in \N$  and $|m|\leq j$
supports the  representation
${\cal D}_j$ of $so(3)$.

Similar results can be obtained for the other algebras $so(2,1)$ and $so(3,2)$.
For instance, for the $so(2,1)$ spanned by $\{K_\pm,K_3\}$,
$\{{\cal L}^m_j(x)\}$
 supports  the irreducible  representation of  the discrete series with  Casimir
${\cal{C}}_{su(1,1)}:= m^2-\frac14$ with $m$ fixed and $j \geq |m|$.

On the other hand,
in general these representations are not faithful because $
{\cal L}_j^m(x)=  {\cal L}_j^{-m}(x) $.
The same difficulty is also present in the spherical harmonic where the  associated Legendre polynomial ${P}_l^{m}$ is related to  ${P}_l^{-m}$. There
the degeneration was removed by introducing an angle variable.  Here we follow the same procedure by
considering  the new functions  
\[
{\cal Z}_j^m(r,\phi) :=  e^{i m \phi}\,
{\cal L}_j^m(r^2)\,,\qquad \phi \in \R\,, \,
 -\pi \le \phi < \pi\,.
\]
Under the change of variable $x\to r^2$ the DE \eqref{defLag2}  becomes
  \[
\left[ \frac{d^2}{dr^2} + \frac{1}{r} \frac{d}{dr}
-\frac{4 m^2}{r^2}- r^2 + 4(j+\frac{1}{2})\right]\,\,{\cal Z}^m_j(r,\phi)
=0\, .
\]
Normalization and orthogonality of the
${\cal Z}_j^m(r,\phi)$ are similar to the ones of $Y_j^m(\theta,\phi)$
\[ \begin{array}{l}\label{OrthNormZ}\displaystyle
\frac{1}{2 \pi} \int_{-\pi}^{\pi}  \, d \phi \int_{0}^{\infty} \;2 r\; dr\;\; {\cal Z}^m_{j}(r, \phi)^*\,
\, {\cal Z}^{m'}_{j'}(r, \phi)\; =\; \delta_{j,j'}\;\; \delta_{m,m'} \;,
\\[0.5cm]
\displaystyle \sum_{j, m}  \;\; {\cal Z}^m_{j}(r, \phi)^*\,\, {\cal Z}^m_{j}(r', \phi')\;
=\;\frac{\pi}{r}\, \delta(r-r')\; \delta(\phi-\phi') \,.
\end{array}\]
This means that the set $\{{\cal Z}_j^m(r, \phi)\}$ is a basis in the  space of  square
integrable functions defined on the plane, $L^2(\R^2)$, like
$\{Y_j^m(\Omega)\}$ is a basis of  $L^2(\S^2)$.

Moreover, with  a convenient  introduction of phases we can define the operators 
$\J_\pm := e^{\pm i \phi}\, J_\pm$ and
$\J_3 := J_3,$ in the finite dimensional space $\{{\cal Z}_j^m(r, \phi)\}$  wih fixed $j$ 
\[\label{actionJ}
\J_\pm\;\; {\cal Z}_j^m(r,\phi)=\ds \sqrt{(j\mp m)(j\pm m+1)} \;\, {\cal Z}_j^{m\pm 1}(r,\phi)\,,
\quad
\J_3\;\; {\cal Z}_j^m(r,\phi)=\ds m \; {\cal Z}_j^{m}(r,\phi) \,,
\]
and analogously for the remaining operators. 
So $\{{\cal Z}_j^m(r,\phi)\}$
support irreducible representations  of $so(3)$, $so(2,1)$ and $so(3,2)$  on the plane
as $\{Y_j^m(\theta, \phi)\}$ are on the sphere. For more details see \cite{olmo15a,iachello83,guerrero06}.

From the physical point of view,  in spite of the analogy with the angular momentum,   $\J_\pm $ and
$\J_3 $  can be  related to a one-dimensional Morse system, where $m$ and $j$ are connected with the potential
\cite{iachello83}.

\section*{Conclusions}

A  relationship between Lie algebras and  square integrable functions has been found.
Indeed we need to restrict ourselves to ${L}^2 (\R^+)$ and 
${L}^2 ( \R^2)$, where $\E$ is identically zero, to obtain  differential representations   of Lie algebras  in the spaces of functions defined in $\R^+$ and 
$\R^2$ .


\section*{Acknowledgements}
This work was partially supported by the Ministerio de Econom\'ia y Competitividad of Spain (Project MTM2014-57129-C2-1-P with EU-FEDER support).


%

\begin{thebibliography}{99.}
%
\bibitem{szego} G. Szeg\"o,  \textit{Orthogonal Polynomials},   (Am. Math. Soc., Providence, 2003), pp. 100-105
%
\bibitem{NITS} F.W.J. Olver, D.W. Lozier, R.F. Boisvert, C.W. Clark, \textit{NIST Handbook of Mathematical Functions}, (Cambridge Univ. Press, New York, 2010)
%
\bibitem{abramowitz} M. Abramowitz, I.A. Stegun, \textit{Handbook of Mathematical Functions}, (Dover, New York, 1972)
%
\bibitem{olmo13} E. Celeghini, M.A. del Olmo, {Ann. Phys.} {\bf 335} (2013) 78-85
%
\bibitem{olmo13a} E. Celeghini, M.A. del Olmo, {Ann. Phys.} {\bf 333} (2013) 90-103
%
\bibitem{olmo15}  E. Celeghini, M.A. del Olmo, M.A.  Velasco,
J. Phys.: Conf. Ser. \textbf { 597} (2015) 012023
%
\bibitem{olmo15a} E. Celeghini, M.A. del Olmo,
    arXiv: 1504.01572  [math-ph]
%
\bibitem{schwinger}
J. Schwinger, in {\it Quantum Theory of Angular Momentum} (L. Biedenharn, E. van Dam,
Eds.), (Academic Press, New York, 1965), pp. 229-279
%
\bibitem{iachello83}
Y. Alhassid, F. G\"ursey, F. Iachello, {Ann. Phys.} {\bf 148} (1983 ) 346-380 
 %
\bibitem{guerrero06} J. Guerrero, V. Aldaya,
J. Phys. A \textbf{39} (2006)  L267-L276.
\end{thebibliography}
\end{document}